\documentclass[a4paper]{article}

\usepackage{INTERSPEECH2019}
\usepackage[table,xcdraw]{xcolor}
\usepackage{breqn}
\usepackage{amsmath}

\title{Non-Autoregressive TTS with Explicit Duration Modelling for Low-Resource Highly Expressive Speech}
\name{Raahil Shah\textsuperscript{*}, Kamil Pokora\textsuperscript{*}, Abdelhamid Ezzerg, Viacheslav Klimkov, Goeric Huybrechts, Bartosz Putrycz, Daniel Korzekwa, Thomas Merritt}
\address{Amazon Text-to-Speech Research}
\email{\{raahshah, kamipoko\}@amazon.com}

\begin{document}

\maketitle	
\begingroup
\renewcommand\thefootnote{*}
\footnotetext{The first two authors have equal contribution.}
\endgroup

\begin{abstract}
Whilst recent neural text-to-speech (TTS) approaches produce high-quality speech, they typically require a large amount of recordings from the target speaker. In previous work  \cite{huybrechts2020lowresource}, a 3-step method was proposed to generate high-quality TTS while greatly reducing the amount of data required for training. 
However, we have observed a ceiling effect in the level of naturalness achievable for highly expressive voices when using this approach.
In this paper, we present a method for building highly expressive TTS voices with as little as 15 minutes of speech data from the target speaker. Compared to the current state-of-the-art approach, our proposed improvements close the gap to recordings by 23.3\% for naturalness of speech and by 16.3\% for speaker similarity.
Further, we match the naturalness and speaker similarity of a Tacotron2-based full-data ($\approx10$ hours) model using only 15 minutes of target speaker data, whereas with 30 minutes or more, we significantly outperform it.
The following improvements are proposed: 1) changing from an autoregressive, attention-based TTS model to a non-autoregressive model replacing attention with an external duration model and 2) an additional Conditional Generative Adversarial Network (cGAN) based fine-tuning step.
\end{abstract}

\section{Introduction}
Recent advancements in the TTS domain have demonstrated highly natural speech generated by neural text-to-speech (NTTS) models  \cite{shen2018natural, skerry2018towards, kalchbrenner2018efficient, oord2018parallel}. However, these models often require large amounts ($\approx10$ hours) of recordings  \cite{chung2019semi} to achieve high levels of naturalness without degradation.

Data collection for TTS is an expensive and time-consuming task. The problem is magnified for highly expressive voices, because it requires higher vocal effort from the voice talent as compared to neutral speech. This amplifies the need for a scalable solution to be able to build highly expressive voices with smaller amounts of data and without substantial cost (i.e. low-resource TTS). 

Previous research around low-resource TTS attempts to address this problem with multi-speaker modelling and transfer learning. Transferring knowledge from full-resource speakers to a low-resource one improves the synthesis quality of the low-resource speaker \cite{chung2019semi, gibiansky2017deep, jia2018transfer, tits2019exploring, latorre2019effect, chen2019end, zhang2020unsupervised}.

Recent work in Huybrechts et al. \cite{huybrechts2020lowresource} brings significant improvements to naturalness by combining multi-speaker modelling with data augmentation for the low-resource speaker. This approach uses a Voice Conversion (VC) model \cite{mohammadi2017overview, hsu2017voice, lorenzo2018voice, kaneko2019cyclegan, karlapati2020copycat} to transform speech from one speaker to sound like speech from another, while preserving the content and prosody of the source speaker. This artificially boosts the training data available for the resource-scarce target speaker by leveraging readily available source speaker data. However, we have observed that this solution does not scale to achieve naturalness on par with a full-data model for more expressive voices than those presented in \cite{huybrechts2020lowresource}. 

To address this limitation, we investigate the most expressive voice in our catalog and propose changes to the model architecture that consistently outperform the approach presented in \cite{huybrechts2020lowresource} and achieve naturalness on par or better than a full-data Tacotron2-based \cite{shen2018natural} model.

First, we propose to switch from a Tacotron2-based (autoregressive) TTS model to a non-autoregressive mel-spectrogram prediction model and to replace the attention mechanism in Tacotron2 with an external duration model. To the authors' knowledge, this work is the first to investigate such NTTS architectures in a reduced data scenario. 
In the literature, so far mainly attention-based or autoregressive models have been explored in the context of expressive low-resource TTS \cite{gibiansky2017deep, xu2020lrspeech, zhang2020adadurian, jia2018transfer}. Such models suffer from stability issues exhibited in synthesised speech, such as babbling, early cut-off, word repetition, and word skipping \cite{he2019robust,zheng-taslp-2019,guo-interspeech-2019,battenberg-icassp-2020}. These problems, attributed to teacher-forcing and attention, are even more prevalent in the reduced data scenario. Recent research in the field \cite{ren2021fastspeech,elias2021parallel,shen-arxiv-2020}, inspired by traditional parametric speech synthesis \cite{Zen_SPSS_SPECOM,ze2013statistical} mitigates these issues by explicitly modelling the durations of phonemes. In addition to improving speech stability, we posit that explicit duration modelling significantly improves the overall naturalness of highly expressive voices by making it easier to model variability in phoneme durations than in the baseline attention-based systems.

Second, we investigate an application of Conditional Generative Adversarial Networks (cGAN) \cite{m014} as an additional fine-tuning step aimed at improving the signal quality of low-resource synthesis. 
The less data we have, the harder it is to maintain good segmental quality and speaker similarity. GANs \cite{GAN2014}, known for generating high quality images, have also been applied in the speech domain to improve the segmental quality of predicted mel-spectrograms \cite{Kaneko2017, Kaneko_2017_Interspeech}. We extend the standard GAN recipe to pass conditioning in addition to the typical mel-spectrogram input to the discriminator. This better informs the discriminator network when making a classification, allowing for more insightful information to flow to the generator.

\section{Proposed Method}\label{sec:method}

As in Huybrechts et al. \cite{huybrechts2020lowresource}, the method presented in this paper is based on three main steps: 1) data augmentation, 2) multi-speaker TTS and 3) fine-tuning. In this work, we also investigate the addition of a fourth step where we fine-tune the model with a cGAN approach to further improve the audio quality. 

Our full proposed low-resource TTS methodology is defined as follows: 
\begin{enumerate}
 \item Train a VC model to augment data for the target speaker.
 \item Train a multi-speaker TTS model using recordings and synthetic data created in Step 1.
 \item Fine-tune the TTS model with the recordings from the target speaker.
 \item Fine-tune the TTS model with the cGAN approach.
\end{enumerate}

The key contribution of this work is the change in TTS architecture from a Tacotron2-style attention-based model to a non-autoregressive acoustic model supported by external durations. The resulting TTS model is comprised of two main components: 1) an acoustic model that predicts mel-spectrogram $\tilde{y}$ from a phoneme sequence $x$, 2) a duration model which assists the acoustic model during inference by providing the duration $\tilde{d}$ of each phoneme. During training, ground truth durations $d$ are used by the acoustic model. As in Huybrechts et al. \cite{huybrechts2020lowresource}, a Parallel WaveNet universal neural vocoder \cite{jiao2021universal} is used to obtain the final speech signal from the generated mel-spectrogram.

\subsection{Voice Conversion Model} \label{sec:copycat}
\begin{figure}[t]
  \centering
  \includegraphics[width=\linewidth]{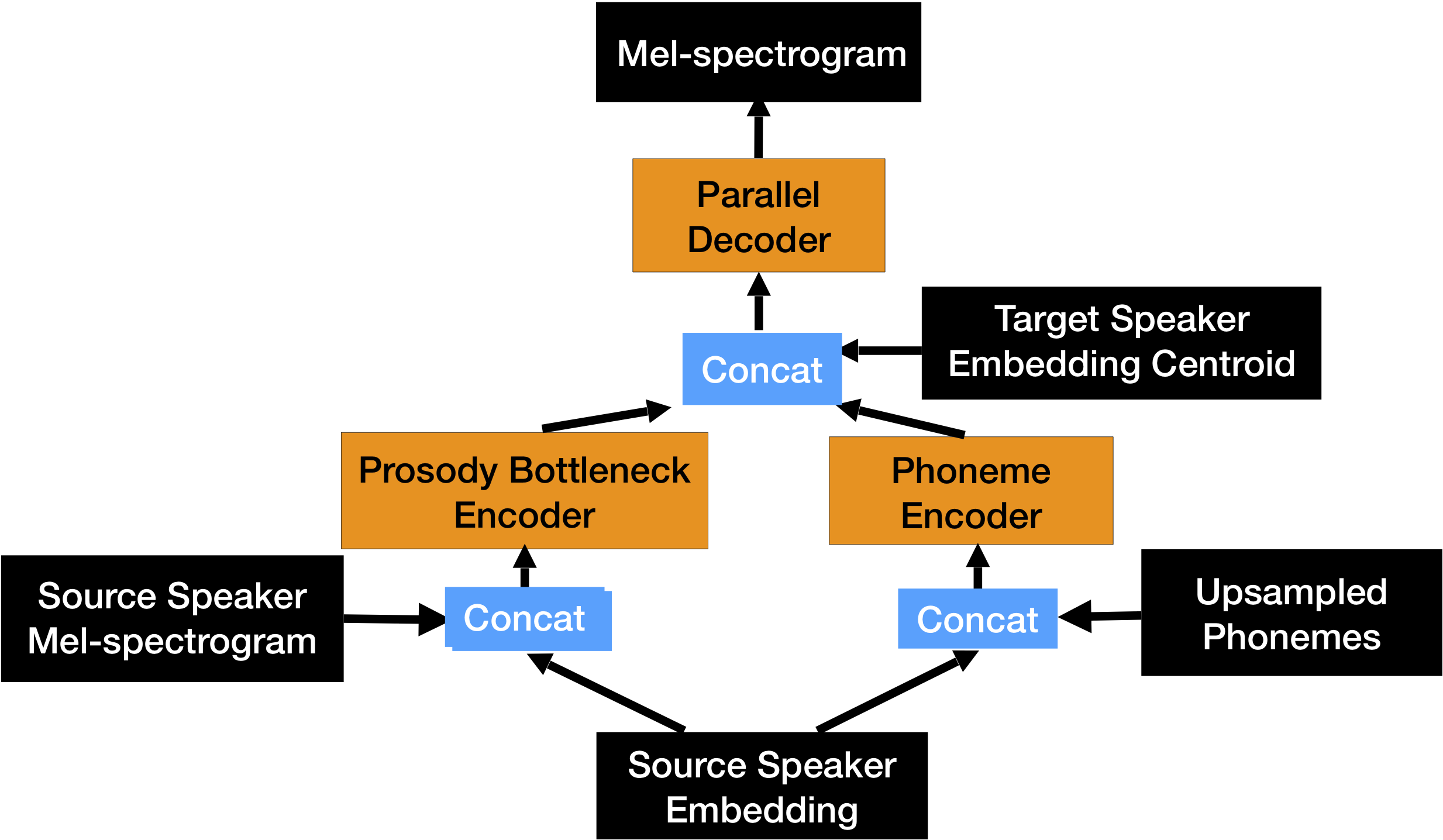}
  \caption{Schematic diagram of the voice conversion model used in Step 1 of the proposed method.}
  \label{fig:copycat}
\end{figure}

A voice conversion model is used to perform data augmentation in Step 1 of the method. This model converts the speaker identity of a source audio to sound as though it was spoken by the target speaker. 

As in Huybrechts et al. \cite{huybrechts2020lowresource}, we use the CopyCat \cite{karlapati2020copycat} architecture for this model which is presented in Figure~\ref{fig:copycat}.  The model consists of: 1) a phoneme encoder that learns latent representations from phonemes, 2) a prosody bottleneck encoder which disentangles prosody from the reference mel-spectrogram and 3) a parallel decoder which generates the mel-spectrogram given the phoneme and prosody bottleneck encoder's outputs, in addition to the target speaker embedding. 

We follow the approach in Huybrechts et al. \cite{huybrechts2020lowresource} to modify the original CopyCat model by concatenating speaker embeddings to the upsampled phonemes before feeding this to the phoneme encoder. This was found to help reduce occurrences of speaker leakage in \cite{huybrechts2020lowresource}.

The VC model was trained with 18 supporting speakers who were recorded in a conversational speaking style, in addition to the target speaker. For the highly expressive target speaker investigated in this paper, fine-tuning of the Copycat model was required to prevent issues with speaker leakage, unlike in Huybrechts et al. \cite{huybrechts2020lowresource}. The model was trained on the data from all speakers for 50 k steps and then only on the target speaker's data for an additional 320 epochs. We hypothesise that this fine-tuning is required because the target speaker's data is much more expressive than that of the supporting speakers.

\subsection{Acoustic Model}

\begin{figure}[t]
  \centering
  \includegraphics[width=\linewidth]{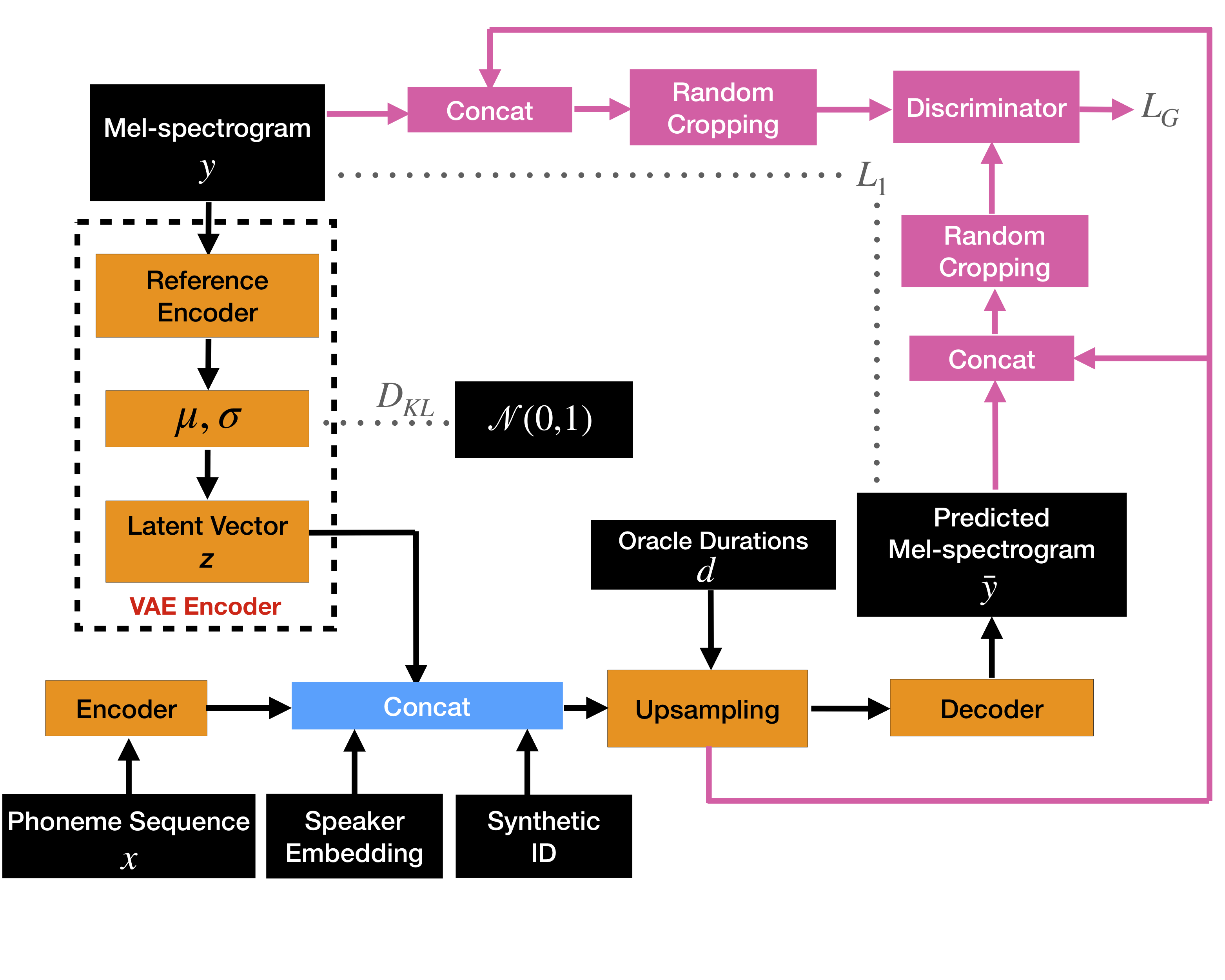}
  \caption{Schematic diagram of the acoustic model used in Steps 2-4 of the proposed method. Components in pink are used only in Step 4 of the method.}
  \label{fig:ttsmodel}
\end{figure}

\begin{figure}[t]
  \centering
  \includegraphics[width=\linewidth]{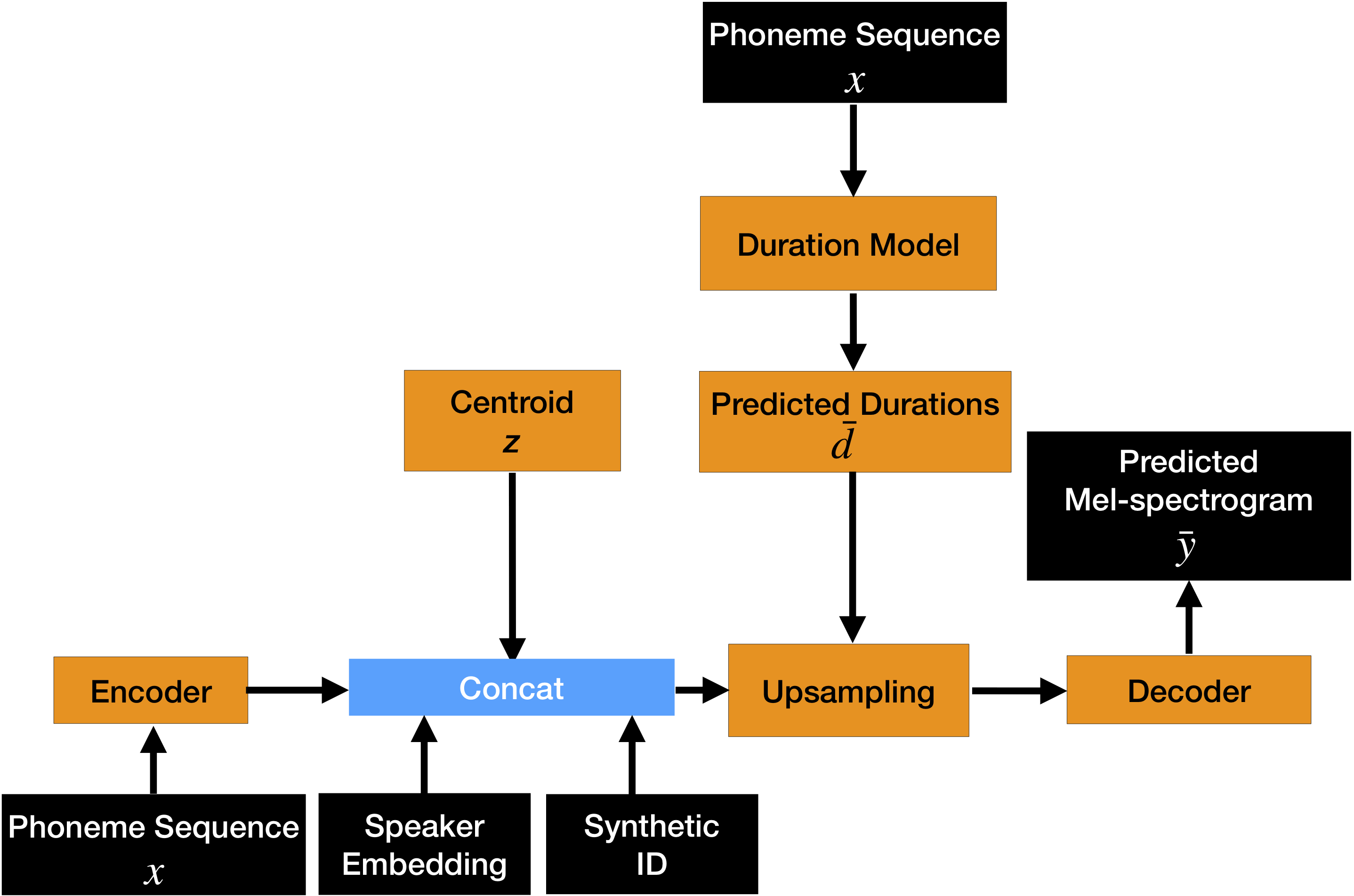}
  \caption{Schematic diagram of the acoustic model during inference.}
  \label{fig:ttsmodelinfer}
\end{figure}

We use an acoustic model in Step 2-4 of the method, as illustrated in Figure~\ref{fig:ttsmodel}. The topology of this model during inference is presented in Figure~\ref{fig:ttsmodelinfer}.

\subsubsection{Encoder}\label{sec:Taco2Encoder}

Our encoder architecture is the same as that presented in Tacotron2 \cite{shen2018natural}. It is comprised of an embedding lookup followed by 3 convolution blocks each with a kernel size of 3. On top of that we apply a single bi-directional LSTM layer with a hidden dimension of 512 and a dropout of 0.1. We pass the phoneme sequence $x$ as input to this encoder, to obtain phoneme embeddings $\tilde{x}$.

\subsubsection{Variational Autoencoder (VAE)}\label{sec:VAEZ}

TTS is a one-to-many problem as the same text can be spoken in many different, yet acceptable, ways. In autoregressive NTTS models, this effect is mitigated both by teacher-forcing as well as by conditioning on the latent acoustic representation obtained from a VAE \cite{kingma2013auto}. In the proposed non-autoregressive architecture, we use only the VAE to pass information which cannot be inferred solely from the input phoneme sequence. 

This encoder takes mel-spectrogram frames as input. It comprises 6 convolution blocks each with a kernel size of 5, followed by one GRU layer with a hidden dimension of 128. We take the last output from the GRU and perform a projection to 128 dimensions, in order to parametrise the posterior distribution. The first half of this output represents $\mu$ while the second half represents $\sigma$. Finally, we sample from the posterior distribution to obtain a final latent representation $z$. At inference time, we use a pre-calculated centroid of $z$s obtained from the available ground truth data for the target speaker.

\subsubsection{Upsampling and Additional Embeddings}

To each phoneme embedding we concatenate: 1) the latent $z$ vector, 2) a speaker embedding obtained from a pre-trained GE2E-based \cite{wan2018generalized} speaker verification model and 3) a one-hot `synthetic ID' flag, indicating whether the data is ground truth or obtained from voice conversion. Then we upsample each phoneme embedding according to ground truth durations $d$ (training-time) or predicted durations $\tilde{d}$ (inference-time). 

Similar to Parallel Tacotron \cite{elias2021parallel}, before passing these upsampled embeddings to the decoder, we provide positional information to indicate the relative position of a frame inside a phoneme. To each embedding we concatenate: 1) a transformer-style positional embedding \cite{vaswani2017attention} indicating the phoneme duration, 2) a transformer-style positional embedding indicating the frame's position inside a phoneme and 3) the fractional progress of the frame in a phoneme.

\subsubsection{Decoder}

\begin{figure}[t]
  \centering
  \includegraphics[scale=0.20]{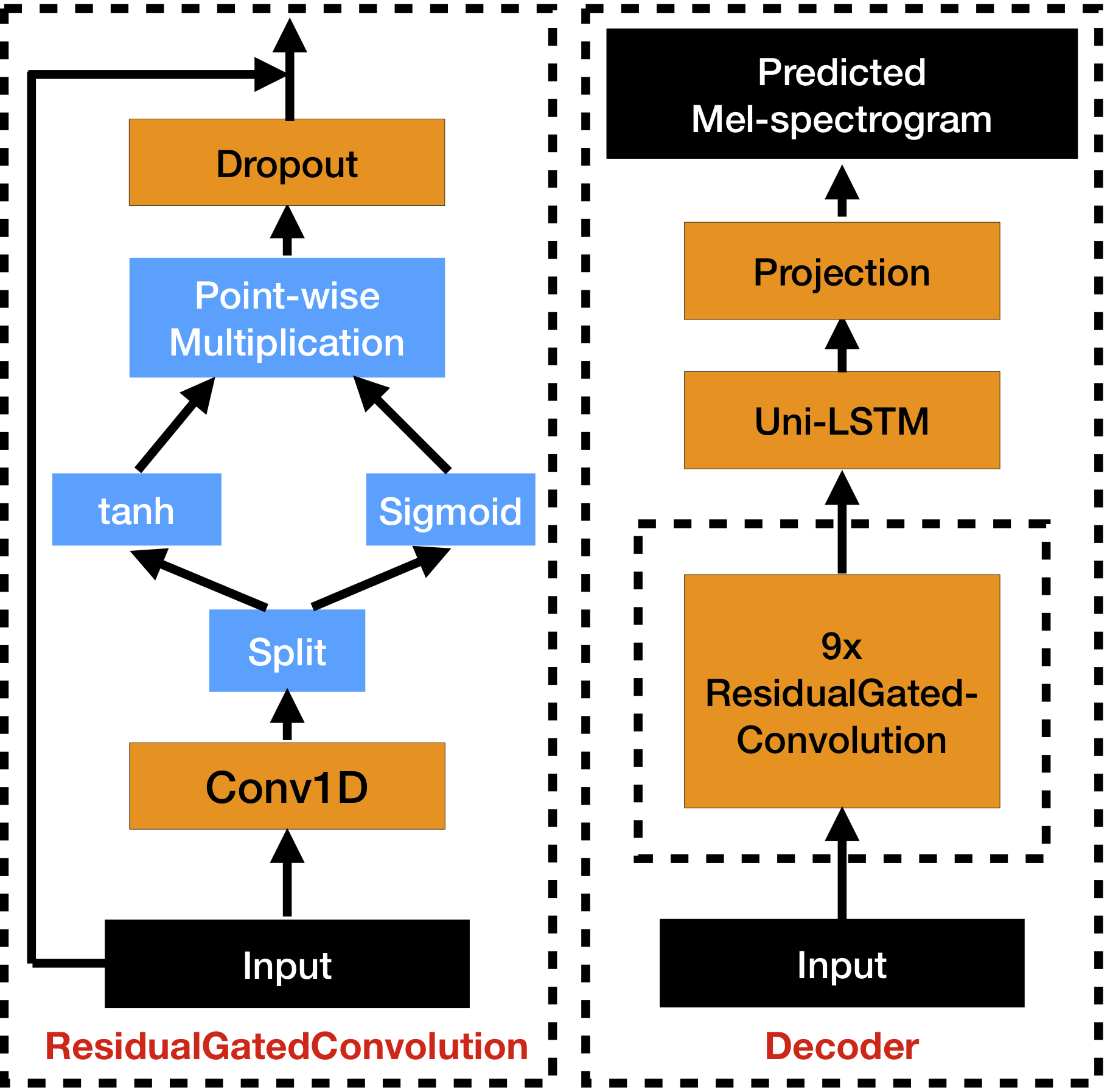}
  \caption{Illustration of a residual gated convolution block and the decoder architecture.}
  \label{fig:decoder}
\end{figure}

The embedding sequence output from the upsampling component is passed as input to the decoder. The modelling task of the decoder was found to require local context in \cite{elias2021parallel}, therefore our decoder is comprised of 9 residual gated convolution layers. Each residual gated convolution block is composed of a 1D-convolution with kernel size 15 and a hidden dimension of 512, followed by a $\tanh$ filter and sigmoid activation gate which are element-wise multiplied and then added to a residual connection after a dropout of 0.1. 

The convolution stack is followed by 2 uni-directional LSTM layers with a hidden dimension of 512 and a dropout of 0.1. Preliminary evaluations showed that this final LSTM stack improves audio quality. A schema of the decoder as well as the residual gated convolution architecture is presented in Figure~\ref{fig:decoder}.

\subsubsection{Conditional GAN Fine-Tuning}

GANs are a well established solution to the problem of `over-smoothing' encountered during the optimisation of L1/L2 loss functions. With mel-spectrogram prediction, this effect manifests as lower brightness and poorer audio quality in the subjective perception of the speech signal. 

Adversarial training of the acoustic model can be utilised as a fine-tuning step to mitigate such degradations \cite{Kaneko2017, Kaneko_2017_Interspeech}. Typically, such an adversarial training involves only the mel-spectrogram being passed as input to the discriminator network. We explore an extension to this setup (cGAN), wherein we condition the discriminator on both acoustic and linguistic information. Additional conditioning allows for more meaningful gradient flow from discriminator to generator, which has been shown to improve adversarial training \cite{m014}. 

The entire acoustic model acts as the generator network. For the discriminator network we used the architecture presented in SAGAN \cite{SAGAN2018}. As input to the discriminator we feed randomly cropped 64 frame chunks of the generated mel-spectrogram, along with the embeddings $\tilde{x}$ of the corresponding phoneme sequence and the latent acoustic information $z$ from the VAE. Cropping was found to be more effective than feeding the whole mel-spectrogram. We hypothesise that this is because the goal of the fine-tuning step is to improve the segmental quality of the final mel-spectrogram, which is a more local, time-invariant task.

\subsubsection{Training Setup}
To train the acoustic model we use the Adam optimiser \cite{kingma2014adam} with $\beta_{1} =0.9$  and $\beta_{2}=0.98$. We use a linear warm-up of the learning rate from 0.1 to 1 for the first 10 k steps, followed by an exponential decay from 10 k steps to 100 k steps with a minimum value of $10^{-5}$.  

In Step 2 of the method, the acoustic model is trained for 500 k steps with a mini-batch size of 32. The model is trained on both ground truth and synthetic data for our target speaker as well as data from supporting speakers, using the following loss function:

\begin{equation}
L_{Train} = L_{1}  + \gamma * D_{KL}
\end{equation}
where $L_{1}$ is the $L_{1}$-distance between predicted and ground truth mel-spectrogram and $D_{KL}$ is the Kullback–Leibler divergence between the VAE posterior distribution and $\mathcal{N}(0,1)$. To avoid the collapse of $D_{KL}$, we used the same KL annealing scheme as presented in \cite{bowman2016generating}.
 
In Step 3 of the method, we fine-tune the model for an additional 30 k training steps, using only ground truth data from the target speaker, still optimising $L_{Train}$. 

Finally, in Step 4 of the method, we freeze all VAE weights and fine-tune the acoustic model with the cGAN setup for an additional 30 k steps, also using only ground truth target speaker data. During this step, the following generator loss ($L_{G}$) and Hinge discriminator loss ($L_{D}$) functions are used:
\begin{equation}
L_{G} = \displaystyle \mathop{\mathbb{E}_{x,y \sim p_{data}}}  [D(y,\tilde{x},V(y)) - D(G(x,y),\tilde{x},V(y))]
\end{equation}
\begin{multline}
L_{D} =  \displaystyle \mathop{\mathbb{E}_{x,y \sim p_{data}}} [\text{ReLU}(1+D(G(x,y),\tilde{x},V(y))) \\ +\text{ReLU}(1-D(y,\tilde{x},V(y)))]
\end{multline}
where $D$ is the discriminator network, $G$ is the generator network (acoustic model) and $V$ denotes the VAE. The discriminator is trained by optimising $L_{D}$, while the acoustic model is fine-tuned by optimising the total loss: 
\begin{equation}
L_{GANFineTune} = L_{1}  +  \alpha * L_{G}
\end{equation}

\subsection{Duration Model}

\begin{figure}[t]
  \centering
  \includegraphics[scale=0.20]{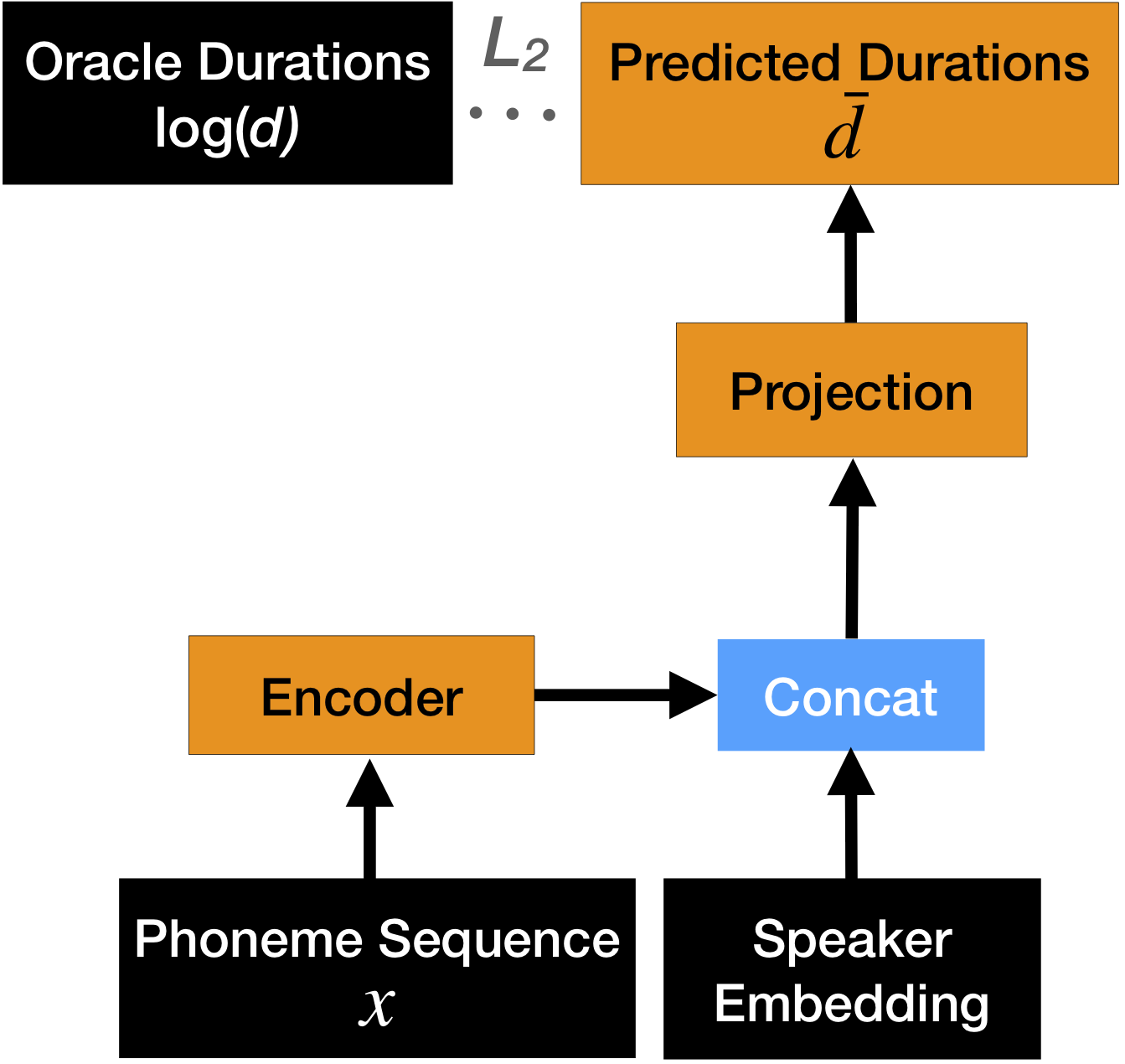}
  \caption{Schematic diagram of the duration model used in Step 2 of the proposed method.}
  \label{fig:duration}
\end{figure}

We train a duration model in Step 2 of the method, whose architecture is presented in Figure~\ref{fig:duration}. We model phoneme durations as the integer number of mel-spectrogram frames corresponding to each phoneme. We assume that ground truth phoneme durations are provided by an external aligner, such as the Gaussian Mixture Model (GMM) based Kaldi Speech Recognition Toolkit \cite{povey2011kaldi} used in our experiments. 

To model the duration sequence, we first pass the phoneme sequence through an encoder and then apply a dense projection to 1 dimension followed by a ReLU activation function. During training, teacher-forcing is used i.e. only ground truth durations are input to the acoustic model, while predicted durations are used only at inference-time.

For the proposed multi-speaker acoustic model using reduced target speaker data, we train the duration model separate from the acoustic model. This model uses a phoneme encoder identical to the one described in Section~\ref{sec:Taco2Encoder}, with the hidden dimension reduced to 256, whose output is concatenated with pre-trained speaker embeddings after they are passed through an affine layer. The training objective for this model is a L2 loss on durations in the $\log$ domain and it is trained for 150 k steps (with a mini-batch size of $32$). We use an identical Adam optimiser configuration as that used for the acoustic model training.

For the full-data anchor models trained on a single-speaker, the embedding sequence used for duration prediction is the concatenation of the phoneme embeddings $\tilde{x}$ and the latent vector $z$ from the acoustic model.  In this setup, in line with the state-of-the-art \cite{ren2021fastspeech, elias2021parallel, shen-arxiv-2020}, the two models are trained jointly, by adding an auxiliary L1 loss between ground truth and predicted durations to the total objective, with a weighting of 0.025.

Preliminary evaluations showed that separately training the duration model in the low-resource multi-speaker scenario performed better than the joint training of acoustic and duration models used for full-data single-speaker models.

\section{Experiments}

\subsection{Data}
For the `highly expressive' target speaker, we selected the voice objectively identified as the most expressive speaker in our internal American English voice catalog. Expressivity was measured as variation along the three axes of frequency, power and durations by analysing respectively the mean and variance of static $\log f0$, $mgc0$ and phoneme duration features and their deltas, for each speaker.

In the `full-data' (FD) setup we used $\approx 10$ hours of recorded speech from the target speaker. In the low-resource aka `data reduction' (DR) scenario, we investigated four different reduced data amounts: 3 hours, 1 hour, 30 minutes and 15 minutes of target speech. To perform data augmentation as detailed in Section~\ref{sec:copycat}, we supplemented the target speaker data in each DR scenario with 4.5 hours of synthetic data converted from the full-data of a single supporting speaker, by a VC model trained using the respective reduced target data amount for that scenario.

For supporting speakers in our multi-speaker models, we used an internal American English dataset comprising 18 speakers recorded in a conversational style. This dataset contained $\approx 65$ hours of speech.

\subsection{Evaluation} 
In each DR scenario, we evaluated our models by conducting MUSHRA tests \cite{itu20031534} on the following two metrics:
\begin{itemize}
	\item{Naturalness} -- ``Please rate the audio samples in terms of their naturalness''.
	\item{Speaker Similarity} -- ``Please listen to the speaker in the reference sample first. Then rate how similar the speakers in each system sound compared to the reference speaker.''
\end{itemize}

Each test was conducted independently, by 20 listeners, each evaluating 61 MUSHRA screens synthesised from a fixed test set of 61 held-out samples. To check for statistical significance, we performed paired t-tests using the Holm-Bonferroni correction method. All statistical differences presented are for $p \leq 0.05$.

\subsection{GAN Fine-Tuning Study}
We conducted a supporting study to investigate improvements from GAN fine-tuning on the naturalness of synthesised speech, demonstrating the impact of Step 4 of the proposed method. In this study, we evaluated the following four systems in a 3 hours DR scenario:
\begin{itemize}
	\item{\textit{(Recordings)}} Ground truth recordings.
	\item{\textit{(DR No-att)}} Baseline without any GAN fine-tuning (i.e. Steps 1-3 of the proposed method).
	\item{\textit{(DR No-att + GAN)}} Candidate system with additional fine-tuning using vanilla GAN (i.e. only mel-spectrogram input to discriminator).
	\item{\textit{(DR No-att + cGAN)}} Candidate system with additional fine-tuning using Conditional GAN (i.e. conditioned on the phoneme sequence, acoustic and prosody information).
\end{itemize}

\begin{table}[h!]
\begin{center}
\scalebox{0.83}{
\begin{tabular}{|
>{\columncolor[HTML]{EFEFEF}}l |c|
>{\columncolor[HTML]{EFEFEF}}l |c|}
\hline
Target Data  & 3 h     \\ \hline
\multicolumn{2}{|c|}{\cellcolor[HTML]{C0C0C0}Naturalness}                                                                                                   \\ \hline
\textit{Recordings}                                      & 88.94                       \\
\textit{DR No-att}                            & 64.45           \\
\textit{DR No-att + GAN}                       &  64.08               \\
\textit{DR No-att +cGAN}                             & 66.57               \\
\hline

\end{tabular}}
\end{center}
\vspace{-0mm}
\caption{\footnotesize{Average MUSHRA scores for naturalness, showing the impact of Conditional GAN fine-tuning.}}
\label{tab:cgan}
\end{table} 

As shown in Table \ref{tab:cgan}, cGAN fine-tuning provides a statistically significant improvement to naturalness when compared to both vanilla GAN fine-tuning and the baseline without GANs. This demonstrates that the addition of conditioning information does indeed appear to help the discriminator make better distinctions of whether a sample is real or fake, which in turn leads to improvements in the samples produced by the generator.

\subsection{Data Reduction Study}
Our primary study investigates the impact of the proposed changes to the model architecture and methodology presented in Huybrechts et al. \cite{huybrechts2020lowresource}, on different amounts of reduced data for the highly expressive target speaker.

In this study, we evaluated the following candidate DR systems: \textit{`DR No-att + cGAN'} and  \textit{`DR No-att'}, i.e. the proposed non-autoregressive, external duration TTS model with and without cGAN fine-tuning respectively. We compared them against a baseline DR system to investigate the ablation of our proposed architectural changes and against full-data anchor systems to investigate the ablation of data amount.

\textit{`DR baseline'} denotes the system presented in Huybrechts et al. \cite{huybrechts2020lowresource} which has been shown to synthesise high quality, expressive voices from as little as 15 minutes of data. The same synthetic data is used in the training of both candidate and baseline DR systems. 

As full-data anchor systems we used: 1) \textit{`FD Tacotron2'} -- a Tacotron2-based TTS model and 2) \textit{`FD No-att'} -- the proposed non-autoregressive TTS model. Both full-data systems used an utterance-level VAE and were single-speaker, i.e. trained on all 10 hours of data from the target speaker.

\begin{table}[h!]
\begin{center}
\scalebox{0.83}{
\begin{tabular}{|l|
>{\columncolor[HTML]{EFEFEF}}l |c|
>{\columncolor[HTML]{EFEFEF}}l |c|
>{\columncolor[HTML]{EFEFEF}}l |c|}
\hline
Target Data  &   \, 3 h   &   1 h   &   30 min   & 15 min       \\ \hline
\multicolumn{5}{|c|}{\cellcolor[HTML]{C0C0C0}Naturalness}                                                                                                   \\ \hline
\textit{Recordings}                                      & 85.70               & 83.61               & 86.39             & 82.30                   \\
\textit{FD No-att}                             & 64.17              & 65.01               & 61.41          &  	69.27            \\
\textit{FD Tacotron2}                       &  58.35               & 58.88              & 55.07            & 63.37                \\
\textit{DR No-att}                             & 62.80               &  \underline{64.96}             & \underline{59.16}              &  	\underline{64.31}               \\
\textit{DR No-att + cGAN }&  \underline{64.94}                  &    \underline{65.29}               & \underline{59.33}             & \underline{64.22}           \\ 
\textit{DR baseline}              			  & 54.40                   & 59.86                 & 51.21             &  	58.73         \\ \hline
\multicolumn{5}{|c|}{\cellcolor[HTML]{C0C0C0}Speaker similarity}                                                                                                 \\ \hline
\textit{Recordings}               & 91.94               & 92.47               & 94.04        & 95.95           \\
\textit{FD No-att}               & 69.90               & 65.21               & 64.85            & 70.37          \\
\textit{FD Tacotron2}               &64.11            & 59.90               & 58.66   & 64.05       \\
\textit{DR No-att}                      &  	69.14           & \underline{66.75}               & \underline{62.94}      & \underline{65.97}         \\
\textit{DR No-att + cGAN}    & \underline{71.54}           & \underline{66.72}                 & \underline{62.95}      & \underline{66.21}            \\
\textit{DR baseline}        & 63.71                 & 60.74                  & 53.58       & 60.43            \\ \hline

\end{tabular}}
\end{center}
\vspace{-0mm}
\caption{\footnotesize{Average MUSHRA scores for naturalness and speaker similarity, showing the performance of proposed method in the context of different amount of data. Note that each column is made up of MUSHRA evaluations for one particular data amount, thus scores are not comparable across different columns. Underlined values signify the best performing system amongst DR systems, up to statistically significant differences.}}
\label{tab:dr}
\end{table}

The results of this study are presented in Table \ref{tab:dr}. They show that in terms of naturalness and speaker similarity, the proposed method, \textit{`DR No-att + cGAN'} significantly outperforms the state-of-the-art approach from Huybrechts et al. \cite{huybrechts2020lowresource} (i.e. \textit{`DR Baseline'}) for every data amount, demonstrating a clear improvement to low-resource TTS. Improvements from the proposed changes to the model architecture are further highlighted by the result that \textit{`DR No-att + cGAN'} significantly outperforms \textit{`FD Tacotron2'} when there is 30 minutes or more of target speaker data (up to 95\% data reduction) and matches it in the 15 minutes scenario. 

Compared to \textit{`FD No-att'}, a full-data model with similar architecture, \textit{`DR No-att + cGAN'} is on par for naturalness while bringing a significant improvement to speaker similarity in the 3 hour scenario and is on par for both metrics in the 1 hour scenario. These results highlight the strength of the proposed 4 step methodology in compensating for the reduction in training data. 

The gap between \textit{`DR No-att'} and \textit{`DR No-att + cGAN'} diminishes as we reduce the data further, suggesting that the audio quality improvements brought about from Step 4 (cGAN fine-tuning) are statistically significant only when a relatively large amount of data (3 hours) is available for the target speaker.


\section{Conclusions}

We proposed improvements to the state-of-the-art low-resource TTS technology presented in Huybrechts et al.  \cite{huybrechts2020lowresource}, addressing its limitations when applied to highly expressive voices. The improvements were to: 1) model architecture, i.e. the switch to a non-autoregressive acoustic model supported by external durations instead of an attention-based, autoregressive Tacotron2 architecture, and 2) methodology, i.e. an additional cGAN fine-tuning step. 

The proposed system significantly outperforms the state-of-the-art in both naturalness and speaker similarity, closing the gap to recordings by 23.3\% and 16.3\% respectively, using as little of 15 minutes of speech from the target speaker. Further, compared to a Tacotron2-based model trained on full-data ($\approx10$ hours of speech), the proposed model is on par with just 15 minutes of target speaker data and significantly improves naturalness and speaker similarity with 30 minutes or more data. Finally, with 3 hours of target speaker data, our proposed architecture with additional cGAN fine-tuning outperforms even a full-data model of a similar architecture.

These contributions demonstrate a robust NTTS method that can build high quality, natural speech from as little as 15 minutes of target speaker data and can scale even to highly expressive voices. Such a method can save substantial cost and time invested in data collection for TTS. 

Future work includes applying the proposed method to more voices that are challenging to model, such as expressive multi-lingual or character voices. Further, we intend to explore fine-grained prosody embeddings to better model and control expressive speech. We also intend to investigate the joint training of acoustic and duration models for the multi-speaker DR scenario, which was found to underperform compared to the separate training approach presented in this paper.

\bibliographystyle{IEEEtran}

\bibliography{template}

\begin{thebibliography}{10}
\providecommand{\url}[1]{#1}
\csname url@samestyle\endcsname
\providecommand{\newblock}{\relax}
\providecommand{\bibinfo}[2]{#2}
\providecommand{\BIBentrySTDinterwordspacing}{\spaceskip=0pt\relax}
\providecommand{\BIBentryALTinterwordstretchfactor}{4}
\providecommand{\BIBentryALTinterwordspacing}{\spaceskip=\fontdimen2\font plus
\BIBentryALTinterwordstretchfactor\fontdimen3\font minus
  \fontdimen4\font\relax}
\providecommand{\BIBforeignlanguage}[2]{{%
\expandafter\ifx\csname l@#1\endcsname\relax
\typeout{** WARNING: IEEEtran.bst: No hyphenation pattern has been}%
\typeout{** loaded for the language `#1'. Using the pattern for}%
\typeout{** the default language instead.}%
\else
\language=\csname l@#1\endcsname
\fi
#2}}
\providecommand{\BIBdecl}{\relax}
\BIBdecl

\bibitem{huybrechts2020lowresource}
G.~Huybrechts, T.~Merritt, G.~Comini, B.~Perz, R.~Shah, and J.~Lorenzo-Trueba,
  ``Low-resource expressive text-to-speech using data augmentation,'' in
  \emph{ICASSP}.\hskip 1em plus 0.5em minus 0.4em\relax IEEE, 2021, pp.
  6593--6597.

\bibitem{shen2018natural}
J.~Shen, R.~Pang, R.~J. Weiss, M.~Schuster, N.~Jaitly, Z.~Yang, Z.~Chen,
  Y.~Zhang, Y.~Wang, R.~Skerrv-Ryan \emph{et~al.}, ``Natural tts synthesis by
  conditioning wavenet on mel spectrogram predictions,'' in
  \emph{ICASSP}.\hskip 1em plus 0.5em minus 0.4em\relax IEEE, 2018, pp.
  4779--4783.

\bibitem{skerry2018towards}
R.~Skerry-Ryan, E.~Battenberg, Y.~Xiao, Y.~Wang, D.~Stanton, J.~Shor, R.~Weiss,
  R.~Clark, and R.~A. Saurous, ``Towards end-to-end prosody transfer for
  expressive speech synthesis with tacotron,'' in \emph{ICML}.\hskip 1em plus
  0.5em minus 0.4em\relax PMLR, 2018, pp. 4693--4702.

\bibitem{kalchbrenner2018efficient}
N.~Kalchbrenner, E.~Elsen, K.~Simonyan, S.~Noury, N.~Casagrande, E.~Lockhart,
  F.~Stimberg, A.~Oord, S.~Dieleman, and K.~Kavukcuoglu, ``Efficient neural
  audio synthesis,'' in \emph{ICML}.\hskip 1em plus 0.5em minus 0.4em\relax
  PMLR, 2018, pp. 2410--2419.

\bibitem{oord2018parallel}
A.~Oord \emph{et~al.}, ``Parallel wavenet: Fast high-fidelity speech
  synthesis,'' in \emph{ICML}.\hskip 1em plus 0.5em minus 0.4em\relax PMLR,
  2018, pp. 3918--3926.

\bibitem{chung2019semi}
Y.-A. Chung \emph{et~al.}, ``Semi-supervised training for improving data
  efficiency in end-to-end speech synthesis,'' in \emph{ICASSP}.\hskip 1em plus
  0.5em minus 0.4em\relax IEEE, 2019, pp. 6940--6944.

\bibitem{gibiansky2017deep}
A.~Gibiansky, S.~{\"O}. Arik, G.~F. Diamos, J.~Miller, K.~Peng, W.~Ping,
  J.~Raiman, and Y.~Zhou, ``Deep voice 2: Multi-speaker neural
  text-to-speech.'' in \emph{NIPS}, 2017.

\bibitem{jia2018transfer}
Y.~Jia \emph{et~al.}, ``Transfer learning from speaker verification to
  multispeaker text-to-speech synthesis,'' in \emph{NIPS}, 2018, pp.
  4480--4490.

\bibitem{tits2019exploring}
N.~Tits \emph{et~al.}, ``Exploring transfer learning for low resource emotional
  tts,'' in \emph{Proceedings of SAI Intelligent Systems Conference}.\hskip 1em
  plus 0.5em minus 0.4em\relax Springer, 2019, pp. 52--60.

\bibitem{latorre2019effect}
J.~Latorre \emph{et~al.}, ``Effect of data reduction on sequence-to-sequence
  neural tts,'' in \emph{ICASSP}.\hskip 1em plus 0.5em minus 0.4em\relax IEEE,
  2019, pp. 7075--7079.

\bibitem{chen2019end}
Y.-J. Chen \emph{et~al.}, ``End-to-end text-to-speech for low-resource
  languages by cross-lingual transfer learning.'' in \emph{Interspeech}, 2019,
  pp. 2075--2079.

\bibitem{zhang2020unsupervised}
H.~Zhang and Y.~Lin, ``{Unsupervised Learning for Sequence-to-Sequence
  Text-to-Speech for Low-Resource Languages},'' in \emph{Interspeech}, 2020,
  pp. 3161--3165.

\bibitem{mohammadi2017overview}
S.~H. Mohammadi \emph{et~al.}, ``An overview of voice conversion systems,''
  \emph{Speech Communication}, vol.~88, pp. 65--82, 2017.

\bibitem{hsu2017voice}
C.-C. Hsu, H.-T. Hwang, Y.-C. Wu, Y.~Tsao, and H.-M. Wang, ``Voice conversion
  from unaligned corpora using variational autoencoding wasserstein generative
  adversarial networks,'' in \emph{Interspeech}, 2017, pp. 3364--3368.

\bibitem{lorenzo2018voice}
J.~Lorenzo-Trueba, J.~Yamagishi, T.~Toda, D.~Saito, F.~Villavicencio,
  T.~Kinnunen, and Z.~Ling, ``The voice conversion challenge 2018: Promoting
  development of parallel and nonparallel methods,'' \emph{arXiv preprint
  arXiv:1804.04262}, 2018.

\bibitem{kaneko2019cyclegan}
T.~Kaneko \emph{et~al.}, ``Cyclegan-vc2: Improved cyclegan-based non-parallel
  voice conversion,'' in \emph{ICASSP}.\hskip 1em plus 0.5em minus 0.4em\relax
  IEEE, 2019, pp. 6820--6824.

\bibitem{karlapati2020copycat}
S.~Karlapati, A.~Moinet, A.~Joly, V.~Klimkov, D.~S'aez-Trigueros, and
  T.~Drugman, ``Copycat: Many-to-many fine-grained prosody transfer for neural
  text-to-speech,'' in \emph{Interspeech}, 2020.

\bibitem{xu2020lrspeech}
J.~Xu \emph{et~al.}, ``Lrspeech: Extremely low-resource speech synthesis and
  recognition,'' in \emph{Proceedings of the 26th ACM SIGKDD International
  Conference on Knowledge Discovery \& Data Mining}, 2020, pp. 2802--2812.

\bibitem{zhang2020adadurian}
Z.~Zhang, Q.~Tian, H.~Lu, L.-H. Chen, and S.~Liu, ``Adadurian: Few-shot
  adaptation for neural text-to-speech with durian,'' \emph{arXiv preprint
  arXiv:2005.05642}, 2020.

\bibitem{he2019robust}
M.~He, Y.~Deng, and L.~He, ``{Robust Sequence-to-Sequence Acoustic Modeling
  with Stepwise Monotonic Attention for Neural TTS},'' in \emph{Interspeech},
  2019, pp. 1293--1297.

\bibitem{zheng-taslp-2019}
Y.~Zheng, J.~Tao, Z.~Wen, and J.~Yi, ``Forward–backward decoding sequence for
  regularizing end-to-end tts,'' \emph{IEEE/ACM Trans. Audio Speech \& Lang.
  Process.}, vol.~27, no.~12, pp. 2067--2079, 2019.

\bibitem{guo-interspeech-2019}
H.~Guo, F.~K. Soong, L.~He, and L.~Xie, ``{A New GAN-Based End-to-End TTS
  Training Algorithm},'' in \emph{Interspeech}, 2019, pp. 1288--1292.

\bibitem{battenberg-icassp-2020}
E.~Battenberg, R.~Skerry-Ryan, S.~Mariooryad, D.~Stanton, D.~Kao, M.~Shannon,
  and T.~Bagby, ``Location-relative attention mechanisms for robust long-form
  speech synthesis,'' in \emph{ICASSP}, 2020, pp. 6194--6198.

\bibitem{ren2021fastspeech}
Y.~Ren, C.~Hu, X.~Tan, T.~Qin, S.~Zhao, Z.~Zhao, and T.-Y. Liu, ``Fastspeech 2:
  Fast and high-quality end-to-end text to speech,'' in \emph{ICLR}, 2021.

\bibitem{elias2021parallel}
I.~Elias, H.~Zen, J.~Shen, Y.~Zhang, Y.~Jia, R.~J. Weiss, and Y.~Wu, ``Parallel
  tacotron: Non-autoregressive and controllable tts,'' in \emph{ICASSP}.\hskip
  1em plus 0.5em minus 0.4em\relax IEEE, 2021, pp. 5709--5713.

\bibitem{shen-arxiv-2020}
J.~Shen, Y.~Jia, M.~Chrzanowski, Y.~Zhang, I.~Elias, H.~Zen, and Y.~Wu,
  ``{Non-Attentive Tacotron: Robust and Controllable Neural TTS Synthesis
  Including Unsupervised Duration Modeling},'' \emph{arXiv:2010.04301}, 2020.

\bibitem{Zen_SPSS_SPECOM}
H.~Zen, K.~Tokuda, and A.~Black, ``{Statistical Parametric Speech Synthesis},''
  \emph{Speech Communication}, vol.~51, no.~11, pp. 1039--1064, 2009.

\bibitem{ze2013statistical}
H.~Ze, A.~Senior, and M.~Schuster, ``Statistical parametric speech synthesis
  using deep neural networks,'' in \emph{ICASSP}.\hskip 1em plus 0.5em minus
  0.4em\relax IEEE, 2013, pp. 7962--7966.

\bibitem{m014}
M.~Mirza and S.~Osindero, ``Conditional generative adversarial nets,''
  \emph{arXiv preprint arXiv:1411.1784}, 2014.

\bibitem{GAN2014}
I.~J. Goodfellow, J.~Pouget-Abadie, M.~Mirza, B.~Xu, D.~Warde-Farley, S.~Ozair,
  A.~Courville, and Y.~Bengio, ``Generative adversarial networks,'' \emph{arXiv
  preprint arXiv:1406.2661}, 2014.

\bibitem{Kaneko2017}
T.~{Kaneko}, H.~{Kameoka}, N.~{Hojo}, Y.~{Ijima}, K.~{Hiramatsu}, and
  K.~{Kashino}, ``Generative adversarial network-based postfilter for
  statistical parametric speech synthesis,'' in \emph{ICASSP}, March 2017, pp.
  4910--4914.

\bibitem{Kaneko_2017_Interspeech}
T.~Kaneko, S.~Takaki, H.~Kameoka, and J.~Yamagishi, ``Generative adversarial
  network-based postfilter for stft spectrograms,'' in \emph{Interspeech},
  2017.

\bibitem{jiao2021universal}
Y.~Jiao, A.~Gabry{\'s}, G.~Tinchev, B.~Putrycz, D.~Korzekwa, and V.~Klimkov,
  ``Universal neural vocoding with parallel wavenet,'' in \emph{ICASSP}.\hskip
  1em plus 0.5em minus 0.4em\relax IEEE, 2021, pp. 6044--6048.

\bibitem{kingma2013auto}
D.~P. Kingma \emph{et~al.}, ``Auto-encoding variational bayes,'' \emph{arXiv
  preprint arXiv:1312.6114}, 2013.

\bibitem{wan2018generalized}
L.~Wan, Q.~Wang, A.~Papir, and I.~L. Moreno, ``Generalized end-to-end loss for
  speaker verification,'' in \emph{ICASSP}.\hskip 1em plus 0.5em minus
  0.4em\relax IEEE, 2018, pp. 4879--4883.

\bibitem{vaswani2017attention}
A.~Vaswani, N.~Shazeer, N.~Parmar, J.~Uszkoreit, L.~Jones, A.~N. Gomez,
  L.~Kaiser, and I.~Polosukhin, ``Attention is all you need,'' \emph{arXiv
  preprint arXiv:1706.03762}, 2017.

\bibitem{SAGAN2018}
H.~Zhang, I.~Goodfellow, D.~Metaxas, and A.~Odena, ``Self-attention generative
  adversarial networks,'' in \emph{International conference on machine
  learning}.\hskip 1em plus 0.5em minus 0.4em\relax PMLR, 2019, pp. 7354--7363.

\bibitem{kingma2014adam}
D.~P. Kingma and J.~Ba, ``Adam: A method for stochastic optimization,''
  \emph{arXiv preprint arXiv:1412.6980}, 2014.

\bibitem{bowman2016generating}
S.~R. Bowman, L.~Vilnis, O.~Vinyals, A.~Dai, R.~Jozefowicz, and S.~Bengio,
  ``Generating sentences from a continuous space,'' in \emph{CoNLL}.\hskip 1em
  plus 0.5em minus 0.4em\relax ACL, 2016, pp. 10--21.

\bibitem{povey2011kaldi}
D.~Povey, A.~Ghoshal, G.~Boulianne, L.~Burget, O.~Glembek, N.~Goel,
  M.~Hannemann, P.~Motlicek, Y.~Qian, P.~Schwarz \emph{et~al.}, ``The kaldi
  speech recognition toolkit,'' in \emph{IEEE 2011 workshop on automatic speech
  recognition and understanding}, no. CONF.\hskip 1em plus 0.5em minus
  0.4em\relax IEEE Signal Processing Society, 2011.

\bibitem{itu20031534}
R.~ITU-R, ``1534-1, method for the subjective assessment of intermediate
  quality levels of coding systems (mushra),”,'' \emph{International
  Telecommunication Union}, 2003.

\end{thebibliography}

\end{document}